\documentclass[11pt,a4paper]{article}
\usepackage{graphicx}
\date{}
\usepackage[latin1]{inputenc}
\usepackage{amsmath}
\usepackage{amsfonts}
\usepackage{amssymb}
\author{ Afranio R. Pereira\\Departamento de F\'isica, Universidade Federal de Vi\c cosa, Vi\c cosa,  36570-000\\Minas Gerais, Brazil.}
\title{\textbf{Heisenberg Spins on a Circular Conical Surface}}
\begin{document}
\maketitle
\begin{center}
\textbf{Abstract}
\end{center}

\indent We investigate classical Heisenberg spins on a conical surface.
The energy and configuration of non-trivial spin distributions are obtained using a non-conventional method based on Einstein theory of gravity in lower dimensions.
Corresponding author: A. R. Pereira\\
e-mail: apereira@ufv.br\\
Tel.: +55-31-3899-2988.\\
Fax: +55-31-3899-2483.\\

\newpage
\indent The study of low-dimensional, artificially structured
materials is becoming increasingly important as we move to an era
of important technological realizations. Physics in two spatial
dimensions has generated a lot of information and surprises.
Moreover, if the system lies on a curved surface, it adds further
richness to the physical phenomena [1-4]. Recently, it has become
possible the fabrication of magnetic films grown on a curved
substrate [5]. Researchers already discovered how to produce
graphitic cones in 1997 [6,7]. In this letter we consider
Heisenberg spins on a conic support. Although we do not know of
physically relevant examples, it should be interesting to
investigate the behavior of topological structures on exotic
surfaces. After all, artificially structured materials are
produced each day leading to new knowledge and technologies.
Besides, the idea that topologically nontrivial excitations arise
in real physical systems had a strong impact on modern physics.
These structures became an object of intensive investigations in
many condensed matter models and particularly they are believed to
play an important role in magnetic systems [8,12]. Our main
interest is to study the configuration and energy of such
topological excitations on
the surface of a cone. \\
\indent As it is well known, the classical two-dimensional static
ferromagnetic and antiferromagnetic materials are well described,
in the continuum limit, by the   non-linear $\sigma$  model. This
theory
on the surface of a curved geometry is [4]\\
\begin{eqnarray}\label{eq:eq1}
H=\frac{J}{2}\int\int_{S}(D_{i}\vec{\psi})(D^{i}\vec{\psi})\sqrt{|g|}d\Omega,\qquad
i=1,2\end{eqnarray} \\
where $D_{i}$ is the covariant derivative, $dS=\sqrt{|g|}d\Omega$
is the surface , $\sqrt{|g|}$ is the determinant of the metric
tensor $g_{ij}$ and the classical spin field $\vec{\psi}$ obeys
the constraint
$\vec{\psi}^{2}=\psi_{1}^{2}+\psi_{2}^{2}+\psi_{3}^{2}=1$. Here,
we consider a non-conventional approach that may simplify the
calculations significantly. Instead of using Hamiltonian (1) with
adequate metric tensor for a cone surface and resolving it, we
apply some results of general relativity. In fact, the conical
geometry has connections with the problem of Einstein gravity in
three space-time $(2+1)$ dimensions. This theory exhibits some
unusual features, which can be deduced from the properties of the
Eintein field equations and the curvature tensor [13]. In regions
which are free of matter, space-time is locally flat when the
cosmological constant is zero [13]. Although the local curvature
in regions free of sources is unaffected by any matter in
space-time, there is still nontrivial global effects. The reason
that these spaces are not completely trivial is because of
nontrivial boundary conditions on the flat coordinates. For
instance, in the case of a massive point-particle sitting at rest
in the origin, we have to remove a wedge out of space-time and
identify opposite points of the wedge. In the static case, the
unique two-dimensional spatial geometry satisfying this
description is the cone [14]. Really, the three-dimensional
space-time around a point-particle of mass $M$ has an interval
given by [13-15]
$ds^{2}=dt^{2}-r^{-8GM}(dr^{2}+r^{2}d\varphi^{2})$, where $G$ is
the gravitational constant and $r,\varphi$  are assumed to be the
polar coordinates. Since the curvature vanishes everywhere except
at the particle position, this interval can be written using local
flat coordinates as $ds^{2}=dt^{2}-d\rho^{2}-\rho^{2}d\phi^{2}$,
where\\
\begin{eqnarray}\label{eq:eq2}
\rho=\frac{r^{\beta}}{\beta},\qquad \phi=\beta\varphi,
\end{eqnarray} \\
and $\beta=1-4GM$. Note that although the situation looks trivial,
the new coordinate $\phi$ ranges from $0$ to $2\pi\beta$,
indicating that there is a deficit angle in space. Thus, it is
always possible to transform to coordinates in which $g_{ij}$
becomes the Minkowski metric, but paying a price, i.e., we will
have to work with multivalued coordinates (or coordinates with
strange boundary conditions). For a static space-time, coordinates
can be chosen for which each of the $t=constant$ spatial sections
are identical. We use these results to study a Heisenberg spin
system on a conic
surface.\\
\indent In the case in which $r,\varphi$  are the polar
coordinates of a point in a two-dimensional plane, the metric
tensor can be written
as\\
\begin{eqnarray}\label{eq:eq3}
g_{ij}=\left[\begin{array}{ccc}1 & 0\\
0 & r^{2}\end{array}\right]
\end{eqnarray} \\
Then, it is easy to write Hamiltonian (1) for spins in a flat
metric. In order to transform this Hamiltonian on a flat space to
the problem of a conical surface, we write first $r$ and $\varphi$
in terms of $\rho$ and $\phi$ respectively, using $\beta=1-\alpha$
for a space with a deficit angle equal to $2\pi\alpha$. It would
imply that the cone surface makes an angle of
$\gamma=\arcsin(\beta)$ with its symmetry z-axis (this axis passes
for a point on the tip of the cone). Thus, Hamiltonian (1) can be
easily written on a conical geometry as
\begin{eqnarray}\label{eq:eq4}
H=\frac{J}{2}\int_{0}^{\infty}\int_{0}^{2\pi\beta}(\partial_{i}\vec{\psi})(\partial^{i}\vec{\psi})\rho
d\rho d\phi,
\end{eqnarray} \\
where $\partial_{i}$ is the gradient written in terms of the
coordinates ($\rho,\phi$). Now, we would like to get nonlinear
excitation solutions with non-zero but finite energy on the
surface of a cone. For the two-dimensional plane ($\beta=1$),
these solutions are the well known Belavin-Polyakov (BP) solitons
[16]. Structurally, these excitations correspond to a mapping of
the spin sphere ($\vec{\psi}^{2}=1$) onto the physical coordinate
plane with a constant field $\vec{\psi}$ far from the soliton
center. Essentially, the plane is compacted into a spherical
surface and $\vec{\psi}$ is a mapping of the sphere to itself
classified by a topological charge $q$ that gives the number of
times $\vec{\psi}$ does this. Hence, there is the Bogomol'nyi
inequality [16] $ H\geq 4\pi J|q|$, where the minimum energy for a
given $q$ is $4\pi J|q|$. In terms of the conformal representation
$w=(\psi_{1}+i\psi_{2})/(1-\psi_{3})$ , which is simply the
stereographic projection of the vector $\vec{\psi}$ onto the
complex plane, the BP-solitons are expressed as
$w_{S}=[(z-z_{0})/R]^{n}$, where $z=x+iy$ and the constants $R,
z_{0}$ appear due to scale and translational symmetry. The energy
of these static structures is finite depending only upon the total
topological charge $|q|=n=1,2,3,\ldots $, and is given by $4\pi
J|q|$; there is no preferred size $R$, position or orientation for
a
fundamental soliton.\\
\indent For the conical geometry ($0<\beta<1$), part of the spin
space sphere is not included in the mapping due to the deficit
angle. Now, the exigency that $\vec{\psi}$ must be constant at
infinity makes that the plane be compacted into a spherical
surface with a hole on it. The area of the hole ($4\pi\alpha$) is
surrounded by two lines (part of geodesics or great circles) drawn
from the top point of the sphere (north pole) to the bottom (south
pole) and separated each other along the equator by an arc length
equal to $2\pi\alpha$. However, the identification of points
across the wedge will close this hole, transforming the space
sphere into a spheroid (something like an oval ball of the
American football) with area $4\pi\beta$. At the same time, the
spin sphere will have to become also a spin spheroid with the same
shape of the space spheroid, due to the boundary conditions used
after to remove a wedge from the physical space, i.e., the spins
will have to obey the relation $\vec{\psi}(0)=\vec{\psi}(2\pi
\beta)$. The surface of the spin spheroid should be written in
terms of a new spin field $ \vec{\chi}=(\beta \psi_{1},\beta
\psi_{2},\psi_{3})$, since this surface is flattened at equator,
which has a radius $\beta$. As a consequence, we will find an
one-to-one mapping from the set of points on the spin spheroid
onto the set points on the cone (stereographic projection).
However, this stereographic projection is difficult to be done
because the spheroid surface does not has a so simple expression.
Then, instead of using the elegant topological arguments as
Belavin and Polyakov did for the flat space, we will solve
explicitly the nonlinear partial differential equations obtained
from Hamiltonian (4). To do this, we will consider explicitly only
excitations with unitary topological charge. The generalization of
the arguments is direct for generic $q$. To solve this Hamiltonian
problem, it is convenient to parametrize the spin field in terms
of two scalar fields $\vec{\psi}=(\sqrt{1-m^{2}}\cos(\Phi),
\sqrt{1-m^{2}}\sin(\Phi), m)$, where $m=\cos(\theta)$, and
$\theta$ and $\Phi$ are the polar and the azimuthal angles
respectively. Then, Hamiltonian (4) is
expressed as\\
\begin{eqnarray}\label{eq:eq5}
H_{c}=\frac{J}{2}\int_{0}^{\infty}\int_{0}^{2\pi\beta}\left[
\frac{(\vec{\nabla}m)^{2}}{1-m^{2}}+(1-m^{2})(\vec{\nabla\Phi})^{2}
\right]\rho d\rho d\phi.
\end{eqnarray} \\
The equations of motion, obtained after variation of the
above Hamiltonian $\delta H_{c}=0$, are given by\\
\begin{eqnarray}\label{eq:eq6}
\nabla^{2}m+\frac{m(\vec{\nabla}m)^{2}}{1-m^{2}}+m(1-m^{2})(\vec{\nabla}\Phi)^{2}=0,
\end{eqnarray}
\begin{eqnarray}\label{eq:eq7}
\nabla^{2}\Phi-\frac{2m\vec{\nabla}m\cdot\vec{\nabla}\Phi}{1-m^{2}}=0.
\end{eqnarray} \\
Clearly, a special solution of Eqs.(6) and (7) is the
soliton $(m_{c},\Phi_{c})$ given by \\
\begin{eqnarray}\label{eq:eq8}
m_{c}=\pm
\frac{\rho^{2}-\rho_{0}^{2}}{\rho^{2}-\rho_{0}^{2}},\qquad
\Phi_{c}=\phi.
\end{eqnarray} \\
where $\rho_{0}$ is a constant. In analogy with the usual case of
a flat space, we associate the signals $+$ and $-$ with
configurations with topological charges equal to $+1$ and $-1$
respectively.\\
\indent Now, to see the configuration of a soliton on the conical
surface more clearly, it would be useful to return to variables
$(r,\varphi)$. Here, since the coordinate $r$ becomes the smallest
distance of a point on the cone surface from the point in which
the cone narrows to, i.e., the origin of the coordinates system or
the tip of the cone, the soliton structure can be written in a
general way as a simple function of $\beta$. In this case, the
identification $\Phi_{c}(0)$ with $\Phi_{c}(2\pi \beta)$ is always
possible for a configuration with radial symmetry. Thus, the
soliton
structure on a conic support is\\
\begin{eqnarray}\label{eq:eq9}
m_{c}=\pm
\frac{r^{2\beta}-R^{2\beta}}{r^{2\beta}+R^{2\beta}},\qquad
\Phi_{c}=\beta \varphi.
\end{eqnarray} \\
where $R=\beta\rho_{0}^{1/\beta}$ is the soliton size, obtained
considering the points on the conical surface in which the spins
are perpendicular to the cone axis (z-axis). These points form a
circumference that divide the cone into two parts with relation to
spin directions: above this line, $m_{c}>0$ ($m_{c}<0$), while
below it, $m_{c}<0$ ($m_{c}>0$). This soliton size can be chosen
arbitrarily since the energy of this structure is
\begin{eqnarray}\label{eq:eq10}
E_{c}=4 \pi J\beta,
\end{eqnarray} \\
and does not depend on  $R$. Then, such an excitation can have
arbitrarily large size, thanks to the scale invariance of the
Hamiltonian (5), and consequently, it can occupy all of space.
Further, this structure has the spin $\vec{\psi}$ pointing in all
different directions as $\vec{r}$ varies, but each cone can
support only one soliton, since there is a preferred position for
its nucleation. In fact, the solutions obtained here cannot move
on the conical surface even for antiferromagnetic systems, which
are, in general, Lorentz invariant. The reasons are simple; first,
the translational symmetry was broken. Second, the Lorentz
transformation is a global one while the cone surface has a
Minkowski metric only locally. Hence these solitons are pinned to
the tip point of the cone. Curiously, pinned non-linear
excitations are also possible in a flat space with vacancies.
Recently, some soliton solutions pinned to a nonmagnetic impurity
were obtained [8-12] and experimentally observed [8,9] in
two-dimensional flat magnetic materials. In this case, part of the
spin space sphere cannot be included in the mapping (around the
impurity), since there is no spin on a vacancy. Essentially, in
these two situations (the flat space with vacancy and the cone),
solitons are pinned to a determined point due to oppositive
causes: in one case, there is a missing spin
while in the other, it is missing space to put spins.\\
\indent The above calculations can be easily generalized to
excitations with $|q|>1$, which leads to a soliton energy equal to
$4\pi J\beta|q|$. Clearly, $q$ gives the number of times the spin
spheroid is traversed as we span the coordinate space as compacted
into another spheroid. Of course, as expected, when $\beta
\rightarrow 1$, $E_{c} \rightarrow 4\pi J|q|$, which is the
BP-soliton energy for solitons in the flat space. It is also
important to note that Eqs. (9) and (10) are explicitly dependent
on the geometry of the support (the parameter $\beta$ or
$\gamma=\arcsin(\beta)$). While a cone does not have a
characteristic length, it has a characteristic angle $\gamma$. The
spin projection along the symmetry axis (z-axis) is extremely
dependent on this angle. In Fig.(1), we plotted $m_{c}$ as a
function of $r$ for $q=\pm 1$, $\beta=1, 1/2$, $1/4$ and $R=5$ (in
units of lattice constants). For small $\beta$ (i.e., a large
deficit angle), $m_{c}(r)$ varies very slowly for $r>R$. Figures
(2) and (3) show schematic representations of the spin
configuration of a magnetic soliton with $q=\pm 1$ on a conical
surface.\\
\indent In summary, we have studied the Heisenberg spins on a
conic support and obtained soliton solutions. It was shown that
solitons can be formed on the tip of a cone. Part of the spin
sphere on the region covered by the deficit angle is not included
in the mapping, which results in a soliton energy equal to $4 \pi
J\beta|q|$. Then, the effect of removing a wedge from the flat
space and identifying points across the wedge is essentially to
change the BP-soliton configuration decreasing its energy. The
results presented here may have relevance for the recently
synthesized carbon nanocones [6] with appropriate magnetic
coatings. If one or more sectors are excised from a single layer
of graphite and the remainder is joined seamlessly, a cone results
[6,17]. By considering the symmetry of a graphite sheet and the
Euler's theorem, it can be shown that only five types of cone can
be made from a continuous sheet of graphite corresponding to the
following values of cone angles $2\gamma=19.2^{o}, 38.9^{o},
60.0^{o}, 84.6^{o}, 112.9^{o}$. These are all synthesized [6]. The
solitons excited on the tips of these magnetically coated carbon
nanocones would have energies given by $E_{c} \simeq 2.1J,4.2J,
6.3J, 8.5J$ and $10.5J$ respectively. Besides the carbon
nanocones, magnetic nanodots can be made of different shapes and
display a variety of magnetic excitations. In the case of a
magnetic dot with conical shape, the above solitons may be
important for the resulting magnetization distributions. Really,
in magnetic nanostructures, solitons can be stabilized by
surface-induced interactions [18]. Another possibility is to study
a spin-polarized electron gas on the conical surface. Such an
electron system was studied on a mesoscopic cylinder [3]. Here,
not only charges but also the spins may develop important role
near the sharp point of the cone. These results are also relevant
to field theories considered on manifolds with nontrivial geometry
and have connection with lower dimensional gravity, mainly
three-dimensional gravity, in which the non-linear $\sigma$ model
is
coupled to Einstein field. \\

 We acknowledge support from CNPq (Brazil).

\newpage
\textbf{Figure Captions}
\\
\\
\\
\textbf{Figure 1}. The z-spin components $m_{c}$ of a soliton with
$q=1$ on a conical surface versus $r$ for $\beta=1$ (solid line),
$\beta=1/2$ (dashed line) and $\beta=1/4$ (dotted line). All these solitons have the same size $R=5$ (in units of lattice constants)\\
\textbf{Figure 2}. Schematic representation of a magnetic
soliton with $q=+1$ on the conic support.\\
\textbf{Figure 3}. Schematic representation of a magnetic soliton
with $q=-1$ on the conic support.

\end{document}